\documentclass[aps,pra,reprint,a4paper,superscriptaddress,showkeys,floatfix]{revtex4-2}
\usepackage[utf8]{inputenc}              

\usepackage[english]{babel}              
                                         
\usepackage[T1]{fontenc}                 

\usepackage{lmodern}                     

\usepackage{microtype}                   
\usepackage{xcolor}                      

\usepackage{graphicx}                    

\usepackage[bf]{subfigure}

\usepackage{amssymb,amsmath,amsfonts}      

\usepackage{mathtools}

\usepackage{braket}                        

\usepackage{tensor}                        

\usepackage[smaller]{cancel}        

\usepackage{upgreek}     
\usepackage{bm}          
\usepackage{enumitem}
\usepackage[pagewise,columnwise]{lineno}
\newcommand{\weq}[1][]{\overset{w_{#1}}{=}}    
\newcommand{\Pb}[2]{\left\{#1,#2\right\}_{P}}    
\newcommand{\Co}[2]{\left[#1,#2\right]_{-}}      
\renewcommand{\bar}[1]{\overline{#1}}
\newcommand{\bs}[1]{\boldsymbol{#1}}

\begin{document}

\title{Effective descriptions of localization in a proper-time parametrized framework}

\date{Version 1.0 as of \today}

\author{E. R. F. Taillebois}
\email{emile.taillebois@ifgoiano.edu.br}
\affiliation{Instituto Federal Goiano - Campus Avançado Ipameri, 75.780-000, Ipameri, Goiás, Brazil}

\author{A. T. Avelar}
\affiliation{Instituto de Física, Universidade Federal de Goiás, 74.690-900, Goiânia,
Goiás, Brazil}

\begin{abstract}
Although the quantization of relativistic systems in a proper-time framework gives new insights concerning the understanding of the so-called localization problem, classical observers cannot be treated as quantum comoving frames and real measurement are typically conceived using an external parameter related to a classical frame. Here, the connection between the proper-time formalism and usual descriptions parametrized by classical observers is obtained by defining a restriction operation that mixes contributions of different values of proper-time. Such a restriction procedure allows us to retrieve the concepts of Newton-Wigner position and Kijowski time of detection in an effective fashion, opening the possibility to interpret the related causalities issues as an apparent phenomena resulting from the time uncertainty that is inherent to every physically acceptable single-particle quantum state.
\end{abstract}



\maketitle

\section{\label{sec:intro}Introduction}

We have recently shown \cite{Taillebois2021a} that the quantization of relativistic systems through a proper-time parametrized formalism is a promising framework in the quest for a solution to the so-called localization problem in relativistic quantum mechanics \cite{Pryce1948, Newton1949, Fleming1965a, Fleming1965, Hegerfeldt1974, Hegerfeldt1980, Hegerfeldt1985}, which is also intimately connected to the issue of the adequate definition of the relativistic spin concept \cite{Peres2002, Peres2004,  Caban2013, Taillebois2013, Bauke2014, Giacomini2019, Taillebois2021}. The proposed formalism is based on a covariant description of localization given by POVMs derived from the components of a closed but non-self-adjoint four-position operator, an approach that leads to the existence of an intrinsic temporal uncertainty for all acceptable single-particle physical states.

The proper-time parametrization has a fundamental character since it does not depend on the properties of an external observer and, therefore, corresponds to an intrinsic description for localization. Physically, this approach implies that the system's time would be observed as classical only if it were possible to define a comoving quantum observer as in \cite{Giacomini2019n}. However, since classical observers cannot be treated as quantum comoving frames, real measurements are typically conceived using an external parameter related to a classical frame, such as the measurement time or the detector's position, instead of the system's proper-time. For this reason, it is necessary to investigate how the proper-time approach connects itself to effective measurements defined using classical observers' parameters.

In the approach developed in \cite{Taillebois2021a}, the quantized versions of the proper-time parametrized four-position variables 
\begin{equation}
Q^{\mu}(\tau) = -\frac{J^{\mu\nu}\Pi_{\nu}}{m^{2}} + \frac{\Pi^{\mu}}{m}\tau, \label{eq:Q}
\end{equation}
were considered, $\Pi_{\mu}$ representing the momentum four-vector, $J^{\mu\nu}$ the angular momentum tensor and $\tau$ the system's proper-time. Despite the invariance of the model with regard to reparametrizations of $\tau$, the choice of the proper-time approach has a fundamental advantage, since it results in a formalism in which time and position are treated on an equal footing and leads to a description of localization that is intrinsic in the sense that it does not depend on classical parameters of the observer itself.

Here, a prescription to connect the proper-time approach to observer dependent measurements is presented. In Section \ref{sc:MotClas}, we give a classical motivation for the procedure to be developed, while in Section \ref{sc:VTF} the associated quantum prescription is defined and applied to instantaneous observation surfaces and to surfaces with one fixed spatial dimension. It is important to note that, due to the intrinsic temporal uncertainty of quantum states, any measurement parametrized by a classical observer corresponds to an effective picture of the system and may lead to apparent paradoxical phenomena. This possibility is evaluated in Section \ref{sc:HEGGG} and concluding remarks are made in Section \ref{sc:disc}.

As in \cite{Taillebois2021a}, the model to be investigated is that of a spinless massive free particle. For completeness, Appendix \ref{app:1} contains a brief account of some results from \cite{Taillebois2021a} that are relevant to the present developments. The natural system of units, with $\hbar=1$ and $c=1$,  is adopted, as well as Minkowski's metric with $(-,+,+,+)$ signature. 

\section{\label{sc:MotClas}Classical Motivation}

In general, classical observers do not have access to the system's proper-time and cannot directly obtain information associated with a fixed value of this parameter. Therefore, measurements performed by classical observers must be defined by means of observation surfaces parametrized by classical variables that establish how the information about $Q^{\mu}(\tau)$ will actually be accessed. In this section, a classical approach to this formalism is presented in order to construct the foundations for its quantum analog.

\subsection{Classical instantaneous slicing}

To begin with, the specific case of instantaneous slicings is developed as a motivation for the general procedure to be presented later. In the instantaneous case, the observer's measurement time is fixed and, consequently, different values of the system's proper-time are accessed for each value of the system's momentum $\Pi_{\mu}$. This amounts to imposing the condition
\begin{equation}
Q^{\mu}(\tau)u_{\mu} = -t, \label{eqRaj1z}
\end{equation}
where $u_{\mu}$ is the four-velocity of the inertial observer and $t$ the classical time associated with the adopted instantaneous observation surface.

Solving \eqref{eqRaj1z} for $\tau$ leads to
\begin{equation}
\tau(t) = - \frac{m}{u^{\sigma}\Pi_{\sigma}}\left[t + Q^{\mu}(0)u_{\mu}\right]. \label{eqTa1}
\end{equation}
Then, replacing \eqref{eqTa1} in \eqref{eq:Q} results in
\begin{equation}
\begin{split}
\tilde{Q}^{\mu}(t)  & \equiv  Q^{\mu}\left(\tau(t)\right) \\ & =  Q^{\mu}(0) - \frac{\Pi^{\mu}}{u_{\sigma}\Pi^{\sigma}}\left[t + Q^{\lambda}(0)u_{\lambda}\right].
\end{split}\nonumber
\end{equation}
The new variables $\tilde{Q}^{\mu}(t)$ return the accessible information concerning $Q^{\mu}(\tau)$ if the observer's measurements are limited by the restriction \eqref{eqRaj1z}. Furthermore, they satisfy the relations
\begin{eqnarray*}
\frac{d \tilde{Q}^{\mu}(t)}{d t} & = & -\frac{\Pi^{\mu}}{u^{\sigma}\Pi_{\sigma}}, \\
\tilde{Q}^{\mu}(t)u_{\mu} & = & -t,
\end{eqnarray*}
that correspond to the expected behavior for a position description parametrized by the observer's time. It is worth noting that the Poisson brackets involving $\tilde{Q}^{\mu}(t)$ are given by
\begin{subequations} \label{eq:ParQ}
\begin{eqnarray}
\Pb{\tilde{Q}^{\mu}(t)}{\tilde{Q}^{\nu}(t)} & = & 0, \\
\Pb{\tilde{Q}^{\mu}(t)}{\Pi^{\nu}} & \weq & \eta^{\mu\nu} - \frac{\Pi^{\mu}u^{\nu}}{u^{\sigma}\Pi_{\sigma}}, \label{eq:ParQ2} \\
\Pb{\tilde{J}^{\mu\nu}}{\tilde{Q}^{\sigma}(t)} & = & 2\Pb{\tilde{Q}^{\sigma}(t)}{\Pi^{[\mu}}\tilde{Q}^{\nu]}(t).
\end{eqnarray}
\end{subequations}
which correspond to the Poisson brackets satisfied by the classical version of the so-called Newton-Wigner position operator \cite{Newton1949,Fleming1965a}. The weak equality $\weq$ in \eqref{eq:ParQ2} simply indicates that the equation is valid over the mass shell constrained surface $\Pi^{\mu}\Pi_{\mu} + m^{2} = 0$.

In order to obtain a way to extend the aforementioned ideas to more general surfaces and, later, to the quantum case, it should be noted that the procedure described above can be performed in an analogous manner using the prescription
\begin{widetext}
\begin{equation}
\begin{aligned}
\left[Q^{\mu}\right]_{Q^{\nu}(\tau)u_{\nu} = -t} & \equiv \hspace{-0.1cm}\underset{-\infty}{\int}^{\infty} \hspace{-0.3cm}d(-u_{\sigma}Q^{\sigma}(\tau))\delta(-u_{\nu}Q^{\nu}(\tau) - t)Q^{\mu}(\tau) \\ & = \underset{-\infty}{\int}^{\infty} \left|u_{\sigma}\frac{d Q^{\sigma}(\tau)}{d \tau}\right|d\tau \delta\left(u_{\nu}Q^{\nu}(\tau)+t\right)Q^{\mu}(\tau) \\
& = \underset{-\infty}{\int}^{\infty}d\tau \delta\left[\tau + \frac{m}{u^{\sigma}\Pi_{\sigma}}\left(t - \frac{u_{\mu}J^{\mu\lambda}\Pi_{\lambda}}{m^{2}}\right)\right]Q^{\mu}(\tau) = \tilde{Q}^{\mu}(t),
\end{aligned}\nonumber
\end{equation}
\end{widetext}
where $\left[Q^{\mu}\right]_{Q^{\nu}(\tau)u_{\nu} = -t}$ indicates the restriction of $Q^{\mu}(\tau)$ to the instantaneous surface $Q^{\nu}(\tau)u_{\nu} = -t$.

\subsection{Arbitrary classical slicing}

An arbitrary observation surface is defined by means of a relationship of the form $f(Q^{\mu}(\tau)) = 0$, while the restriction of a physical quantity $A(\tau)$ to that surface can be obtained by means of the prescription
\begin{equation}
\left[A\right]_{f(Q^{\mu}(\tau)) = 0} \equiv \int_{-\infty}^{\infty}\hspace{-10pt}d\tau \Bigg|\frac{df(Q^{\mu}(\tau))}{d\tau}\Bigg|\delta\left[f(Q^{\mu}(\tau))\right]A(\tau). \label{eq:Marolf}
\end{equation}
The construction of physical quantities using relations similar to \eqref{eq:Marolf} appears in \cite{Marolf1995(1),Marolf1995(2)}. However, unlike what is done here, in those works these operations were used as a tool for the construction of observables invariant by temporal reparametrization in the classical non-calibrated space and in the quantum auxiliary Hilbert space.

From \eqref{eq:Marolf} it is immediate to realize that the classical version of the restriction operation satisfies the following properties: 
\begin{enumerate}[label=\roman*)] 
\item If $A(\tau) \equiv A$, then $\left[A\right]_{f(Q^{\mu}(\tau)) = 0} = A$; \label{en:prop1}
\item If $A(\tau) = f(Q^{\mu}(\tau))$, then $[f]_{f(Q^{\mu}(\tau)) = 0} = 0$. \label{en:prop2}
\end{enumerate}\vspace{\jot}
In what follows, the quantum formalism for restriction measurement surfaces is constructed in such a way as to recover these properties.

\section{\label{sc:VTF}Quantum restrictions}

The restriction to observation surfaces becomes more intricate in the quantum case due to the momentum superposition of general quantum states. Therefore, fixing a classical observer parameter does not set a single value for the system's proper time as was the case in \eqref{eqTa1}, since this relationship is momentum dependent and will be affected by the quantum superposition of momenta. Thus, it is necessary to define a quantity that describes only the information of $Q^{\mu}(\tau)$ that is actually accessible to an observer in a given measurement procedure.

The quantum restriction process to be developed is designed to restrict operators to a surface specified by a relationship of the form $\hat{f}(\hat{Q}^{\mu}(\tau)) \mapsto 0$, where the arrow notation indicates that the restraining process will map the operator $\hat{f}(\hat{Q}^{\mu}(\tau))$ into a classical number. This process has ambiguities in its definition due to the ordering of terms in \eqref{eq:Marolf}. To avoid these ambiguities, the classical properties {\ref{en:prop1} and \ref{en:prop2} will be used, their quantum analogs being given by the following conditions:
\begin{enumerate}[label=\roman*)] 
\item If $\hat{A}(\tau) \equiv \hat{A}$ and $\Co{\hat{A}}{\hat{f}(\hat{Q}^{\mu}(\tau))} = 0$, then $[\hat{A}]_{\hat{f}(\hat{Q}^{\mu}(\tau)) \mapsto 0} = \hat{A}$;
\item If $\hat{A}(\tau) = \hat{f}(\hat{Q}^{\mu}(\tau))$, then $[\hat{f}]_{\hat{f}(\hat{Q}^{\mu}(\tau)) \mapsto 0} = 0$.
\end{enumerate}

Of the three possible arrangements for the quantization of \eqref{eq:Marolf}, the only one that meets the above conditions is given by
\begin{widetext}
\begin{equation}
\left[\hat{A}\right]_{\hat{f}(\hat{Q}^{\mu}(\tau)) \mapsto 0} = \frac{1}{2\uppi}\int_{-\infty}^{\infty}d\tau \Bigg|\frac{d\hat{f}(\hat{Q}^{\mu}(\tau))}{d\tau}\Bigg| : \left(\int_{-\infty}^{\infty} d\alpha e^{i\alpha\hat{f}(\hat{Q}^{\mu}(\tau))} : \hat{A}(\tau)\right), \label{eq:QuantRest}
\end{equation}
\end{widetext}
where the Dirac delta distribution was replaced by its integral form. To verify that, it is enough to expand the operator $\hat{A}(\tau)$ in terms of a complete basis that diagonalizes it. The other symmetrizations will not present the required properties due to commutation problems involving the term $|d\hat{f}(\hat{Q}^{\mu}(\tau))/d\tau|$.

Next, the prescription in \eqref{eq:QuantRest} will be used to restrict the operators associated with the variables $Q^{\mu}(\tau)$ to the instantaneous surfaces introduced in the previous section. Subsequently, the method will also be applied to surfaces with a fixed spatial dimension. The single-particle operators associated with the POVM measurements introduced in \cite{Taillebois2021a} will be used in both cases and, therefore, for now on we will be using the notation $\underline{\hat{Q}^{\mu;\xi}_{phys}(\tau)}$ instead of $\hat{Q}^{\mu}(\tau)$, indicating the closed nature of the operators. The index $\xi = \pm$ indicates the subspace $\mathcal{H}_{phys}^{\xi}$ of well-defined energy sign over which the operators are defined.

\subsection{Instantaneous surfaces of observation}

An instantaneous observation surface is defined by setting the measurement time of a classical inertial observer, i.e. this surface is parametrized by the time $t$ measured by a classical clock associated with the observer. The associated restriction relation is given by $\hat{f}_{t}(\underline{\hat{Q}^{\mu;\xi}_{phys}(\tau)})\mapsto 0$, where
\begin{equation}
\hat{f}_{t}(\underline{\hat{Q}^{\mu;\xi}_{phys}(\tau)}) = \underline{\hat{Q}^{\mu;\xi}_{phys}(\tau)}u_{u} + t = -\underline{\hat{Q}^{0;\xi}_{phys}(\tau)} + t \label{eq:SupInst}
\end{equation}
and $u^{\mu}$ is the four-velocity of the inertial observer with respect to which the surface is defined.

By replacing \eqref{eq:SupInst} in \eqref{eq:QuantRest}, one has that the restriction of the operators $\underline{\hat{Q}^{j;\xi}_{phys}(\tau)}$ to instantaneous surfaces will be given by
\begin{widetext}
\begin{equation}
\left[\underline{\hat{Q}^{j,\xi}_{phys}(\tau)}\right]_{-\underline{\hat{Q}^{0;\xi}_{phys}(\tau)} + t \mapsto 0} = \frac{1}{2\uppi}\int_{-\infty}^{\infty}d\tau \frac{\hat{E}_{\bs{\pi}}}{m} : \left(\int_{-\infty}^{\infty} d\alpha e^{i\alpha\left(-\underline{\hat{Q}^{0;\xi}_{phys}(\tau)} + t\right)} : \underline{\hat{Q}^{j,\xi}_{phys}(\tau)}\right), \label{eq:RRRES}
\end{equation}
where $\hat{E}_{\bs{\pi}} = \sqrt{\|\mathbf{\hat{\Pi}}\|^2 + m^2}$. On the other hand, $\left[\underline{\hat{Q}^{0,\xi}_{phys}(\tau)}\right]_{-\underline{\hat{Q}^{0;\xi}_{phys}(\tau)} + t \mapsto 0}= t\hat{I}^{\xi}_{phys}$, which agrees with property ii) presented in Section \ref{sc:VTF}. From this result it can be seen that the restriction process breaks the symmetry existing between time and position in the proper-time formalism, since the time component of the four-position becomes a classic parameter after the instantaneous restriction.

Using the completeness relationships \eqref{eq:IdentP1} and \eqref{eq:IdentP2} the restriction of the operators $\underline{\hat{Q}^{j;\xi}_{phys}(\tau)}$ to a surface of the instantaneous type will be given, in the subspaces with well-defined energy sign, by the acting rule
\begin{equation}
\braket{\psi|\left[\underline{\hat{Q}^{j,\xi}_{phys}(\tau)}\right]_{-\underline{\hat{Q}^{0;\xi}_{phys}(\tau)} + t \mapsto 0}|\psi} = \int_{\mathbb{R}^{3}}d\mu(\bs{\pi})\bar{\psi_{\xi}(\bs{\pi})} \left[\xi i \left(\frac{\partial}{\partial \pi_{j}} - \frac{\pi^{j}}{2E_{\bs{\pi}}^2}\right) + \frac{\pi^{j}}{E_{\bs{\pi}}}t\right]\psi_{\xi}(\bs{\pi}), \label{eq:RegAt}
\end{equation}
\end{widetext}
where $\pi^{\mu}$, with $|\pi^{0}| = E_{\bs{\pi}}$, indicates the eigenvalues of $\hat{\Pi}^{\mu}$. The operation obtained in \eqref{eq:RegAt} is associated to the well-known Newton-Wigner position operator $\hat{\bs{X}}^{\xi}_{NW}(t)$ \cite{Newton1949}, which is the most common relativistic position definition in the relativistic quantum mechanics literature. This operator satisfies commutation relations analogous to the classical Poisson brackets in \eqref{eq:ParQ} and can be interpreted as a quantum analog of the classical variables $\tilde{Q}^{\mu}(t)$.

The components of $\hat{\bs{X}}^{\xi}_{NW}(t)$ commute with each other and form a complete set of commuting observables with spectrum $\mathbf{x} \in \mathbb{R}^3$ associated to the generalized eigenfunctions
\begin{equation}
\psi^{\mathbf{x};\xi}_{NW}(\bs{\pi};t) = \frac{1}{(2\uppi)^{3/2}}\sqrt{\frac{E_{\bs{\pi}}}{m}}e^{i(\xi E_{\bs{\pi}}t-\bs{\pi}\cdot\mathbf{x})}, \nonumber
\end{equation}
which form an improper orthogonal basis for the physical Hilbert space $\mathcal{H}^{\xi}_{phys}$. Although the components of $\hat{X}^{\xi;j}_{NW}(t) \equiv [\underline{\hat{Q}^{j,\xi}_{phys}(\tau)}]_{-\underline{\hat{Q}^{0;\xi}_{phys}(\tau)} + t \mapsto 0} $ satisfy a number of requirements commonly taken as fundamentals to define a position operator in an instant of time $t$ \cite{Newton1949}, the self-adjoint character of $\hat{\bs{X}}^{\xi}_{NW}(t)$ implies that its eigenfunctions are associated with a strict notion of localization, resulting, according to Hegerfeldt's Theorem \cite{Hegerfeldt1974,Hegerfeldt1980,Hegerfeldt1985}, in a violation of relativistic causality for an evolution in $t$ \cite{Ruijsenaars1981}. A discussion about this apparent violation based on the single-particle proper-time parametrized formalism is presented in Section \ref{sc:HEGGG}.

\subsection{Surfaces with one fixed spacial dimension}

Previously it was shown that the restriction of operators $\underline{\hat{Q}^{\mu;\xi}_{phys}(\tau)}$ to a instantaneous observation surface allows to recover the usual Newton-Wigner definition of position for spinless massive systems. This allows to interpret the results associated to this operator in terms of a single-particle formalism parametrized by the system's proper-time, introducing a new possibility of investigation regarding the problem of the localization of relativistic quantum systems. However, to validate this approach, it is important to question whether the prescription introduced in \eqref{eq:QuantRest} produces satisfactory results for other types of restrictions, i.e., it should be verified that the result obtained for $\hat{\bs{X}}_{NW}^{\xi}(t)$ is not only due to coincidence. To investigate that, in this section the procedure introduced in \eqref{eq:QuantRest} will be applied to the case of a three-dimensional surface defined by setting a spatial coordinate.

The restriction surface to be considered here is defined by means of the relation $\hat{f}_{z}(\underline{\hat{Q}^{\mu;\xi}_{phys}(\tau)}) \mapsto 0$, where
\begin{equation}
\hat{f}_{z}(\underline{\hat{Q}^{\mu;\xi}_{phys}(\tau)}) = \underline{\hat{Q}^{3;\xi}_{phys}(\tau)} - z. \label{eq:SupKij}
\end{equation}
One can interpret this observation surface as a fixed detector on the $z$ axis that may record the information about the system's detection position on the $x\times y$ plane and the corresponding detection time. Thus, if the proposed restriction procedure and the interpretation given to the operators  $\underline{\hat{Q}^{\mu;\xi}_{phys}(\tau)}$ are consistent, it is expected that the operators obtained via \eqref{eq:SupKij} will be related to the so-called Kijowski distribution for detection times \cite{Kijowski1974}.

By replacing \eqref{eq:SupKij} in \eqref{eq:QuantRest} one has that the restrictions of the operators $\underline{\hat{Q}^{\mu;\xi}_{phys}(\tau)}$ will be given by
\begin{widetext}
\begin{equation}
\left[\underline{\hat{Q}^{\mu,\xi}_{phys}(\tau)}\right]_{\underline{\hat{Q}^{3;\xi}_{phys}(\tau)} - z \mapsto 0} = \frac{1}{2\uppi}\int_{-\infty}^{\infty}d\tau \Bigg|\frac{\hat{\Pi}^{3}}{m}\Bigg| : \left(\int_{-\infty}^{\infty} d\alpha e^{i\alpha\left(\underline{\hat{Q}^{3;\xi}_{phys}(\tau)} - z\right)} : \underline{\hat{Q}^{\mu,\xi}_{phys}(\tau)}\right), \label{eq:RRRES1}
\end{equation}
with $[\underline{\hat{Q}^{3,\xi}_{phys}(\tau)}]_{\underline{\hat{Q}^{3;\xi}_{phys}(\tau)} - z \mapsto 0}= z\hat{I}^{\xi}_{phys}$, as expected from property ii) of Section \ref{sc:VTF}. Unlike in the previous case, here it is the operator associated with the position on the $z$ axis that assumes the role of classic parameter after the restriction procedure, while time and position on the $x$ and $y$ axes remain as operators.

Using the completeness relationships \eqref{eq:IdentP1} and \eqref{eq:IdentP2}, the restrictions \eqref{eq:RRRES1} are given, in the subspaces with well-defined energy sign, by the acting rules
\begin{subequations}
\begin{eqnarray}
\braket{\psi|\left[\underline{\hat{Q}^{0,\xi}_{phys}(\tau)}\right]_{\underline{\hat{Q}^{3;\xi}_{phys}(\tau)} - z \mapsto 0}|\psi} & = & \int_{\mathbb{R}^{3}}d\mu(\bs{\pi})\bar{\psi_{\xi}(\bs{\pi})} \frac{E_{\bs{\pi}}}{m}\left[\xi i \left(-\frac{\partial}{\partial \pi_{3}} + \frac{1}{2\pi^3}\right) + z\right]\psi_{\xi}(\bs{\pi}), \label{eq:RegAt1} \\
\braket{\psi|\left[\underline{\hat{Q}^{j,\xi}_{phys}(\tau)}\right]_{\underline{\hat{Q}^{3;\xi}_{phys}(\tau)} - z \mapsto 0}|\psi} & = & \int_{\mathbb{R}^{3}}d\mu(\bs{\pi})\bar{\psi_{\xi}(\bs{\pi})} \left[\xi i \left(\frac{\partial}{\partial \pi_{j}} - \frac{\pi^{j}}{\pi^{3}}\frac{\partial}{\partial \pi_{3}} + \frac{\pi^{j}}{2(\pi^{3})^2}\right) + \frac{\pi^{j}}{\pi^3}z\right]\psi_{\xi}(\bs{\pi}), \label{eq:RegAt2}
\end{eqnarray}
\end{subequations}
with $j=1,2$. It is interesting to note that the operations given in \eqref{eq:RegAt1} and \eqref{eq:RegAt2} commute with each other and therefore the corresponding operators form a complete set of commuting observables. To demonstrate that these operations are in conformity with the relativistic definitions of detection time and position proposed by Kijowski \cite{Kijowski1974}, it is necessary to make some changes in the notation of the equations \eqref{eq:RegAt1} and \eqref{eq:RegAt2}. These changes will be made into the positive energy subspace, which corresponds to the case investigated by Kijowski in \cite{Kijowski1974}. 

Following the notation adopted in \cite{Kijowski1974}, one defines by $\psi_{z}(\bs{\pi}) = \sqrt{\frac{m}{|\pi^{3}|}}\psi_{+}(\bs{\pi})$ the state restricted to the observation surface. With this definition, equations \eqref{eq:RegAt1} and \eqref{eq:RegAt2} can be rewritten as
\begin{eqnarray*}
\braket{\psi|\left[\underline{\hat{Q}^{0,+}_{phys}(\tau)}\right]_{\underline{\hat{Q}^{3;+}_{phys}(\tau)} - z \mapsto 0}|\psi} & = & \underset{\mathbb{R}^{3}}{\int}d\mu(\bs{\pi})\bar{\psi_{z}(\bs{\pi})} \mathrm{sign}(\pi^{3})\frac{E_{\bs{\pi}}}{m}\left( -i\frac{\partial}{\partial \pi_{3}}  + z\right)\psi_{z}(\bs{\pi}),  \\
\braket{\psi|\left[\underline{\hat{Q}^{j,+}_{phys}(\tau)}\right]_{\underline{\hat{Q}^{3;+}_{phys}(\tau)} - z \mapsto 0}|\psi} & = & \underset{\mathbb{R}^{3}}{\int}d\mu(\bs{\pi})\bar{\psi_{z}(\bs{\pi})} \frac{|\pi^{3}|}{m}\left[i \left(\frac{\partial}{\partial \pi_{j}} - \frac{\pi^{j}}{\pi^{3}}\frac{\partial}{\partial \pi_{3}}\right) + \frac{\pi^{j}}{\pi^3}z\right]\psi_{z}(\bs{\pi}).
\end{eqnarray*}
Finally, performing the change of variables $\pi^{3} \mapsto s = \mathrm{sign}(\pi^{3})E_{\bs{\pi}}$, with $s\in \mathbb{R}\setminus(-m,m)$, one has that the above relations result in the rules given by
\begin{eqnarray*}
\braket{\psi|\left[\underline{\hat{Q}^{0,+}_{phys}(\tau)}\right]_{\underline{\hat{Q}^{3;+}_{phys}(\tau)} - z \mapsto 0}|\psi} & = & \hspace{-0.5cm}\underset{\mathbb{R}\setminus(-m,m)}{\int}\hspace{-0.5cm}ds\int_{-\infty}^{\infty}\int_{-\infty}^{\infty}\hspace{-0.5cm}d\pi^{1}d\pi^{2}\bar{\psi_{z}(\pi^{1},\pi^{2},s)}\left(-i\mathrm{sign}(s)\frac{\partial}{\partial s} + \frac{sz}{\sqrt{s^2 - \rho_{\pi}^2 - m^2}}\right)\psi_{z}(\pi^1,\pi^2,s), \\
\braket{\psi|\left[\underline{\hat{Q}^{j,+}_{phys}(\tau)}\right]_{\underline{\hat{Q}^{3;+}_{phys}(\tau)} - z \mapsto 0}|\psi} & = & \hspace{-0.5cm}\underset{\mathbb{R}\setminus(-m,m)}{\int}\hspace{-0.5cm}ds\int_{-\infty}^{\infty}\int_{-\infty}^{\infty}\hspace{-0.5cm}d\pi^{1}d\pi^{2}\bar{\psi_{z}(\pi^{1},\pi^{2},s)}\left(i\frac{\partial}{\partial \pi_{j}} + \frac{\pi^{1}z}{\mathrm{sign}(s)\sqrt{s^2 - \rho_{\pi}^2 - m^2}}\right)\psi_{z}(\pi^1,\pi^2,s),
\end{eqnarray*}
\end{widetext}
where $\rho_{\pi}^{2} = (\pi^1)^2 + (\pi^2)^2$ and $j=1,2$. For $z=0$, the relations above coincide with those obtained in \cite{Kijowski1974} and, therefore, the operations found through the restriction process \eqref{eq:QuantRest} correspond to generalizations of the detection time and position operators of Kijowski for the case in which the observation plane is set at any value of $z\in\mathbb{R}$. This result supports the approach proposed here in terms of the restriction of the four-position operators $\underline{\hat{Q}^{\mu;\xi}_{phys}(\tau)}$ to observation surfaces parametrized by classical observer properties.

In the above description, terms with $s > m$ correspond to detections from the left in the observation plane, while terms with $s < m$ correspond to detections from the right. The procedure adopted in \cite{Kijowski1974} for the construction of detection time operators allows these contributions to be combined in two distinct ways, one that has no self-adjoint extensions and the other that has such extensions. It's interesting to note that the operator $[\underline{\hat{Q}^{0,+}_{phys}(\tau)}]_{\underline{\hat{Q}^{3;+}_{phys}(\tau)} - z \mapsto 0}$ obtained through the restriction process \eqref{eq:QuantRest} corresponds to the case that has no self-adjoint extensions and is therefore associated with a POVM-type measurement, a result that is in accordance with more recent descriptions of the detection time distributions \cite{Egusquiza2008}.

\section{\label{sc:HEGGG}Apparent causality issues}

The causality violation associated with the Newton-Wigner eigenfunctions usually justifies the impossibility of constructing a satisfactory definition of relativistic quantum localization. However, assuming that the fundamental concept of system position is given by the POVMs defined from $\underline{\hat{Q}^{j;\xi}_{phys}(\tau)}$ and interpreting the operators $\hat{X}^{\xi;j}_{NW}(t)$ as effective position definitions resulting from the restriction of $\underline{\hat{Q}^{j;\xi}_{phys}(\tau)}$, the possibility that this violation is an apparent phenomenon resulting from the restriction process cannot be excluded.

Since the single-particle proper-time approach assumes that the fundamental definition of position is given by the POVMs associated with $\underline{\hat{Q}^{\mu;\xi}_{phys}(\tau)}$, it should be possible to write the components of $\hat{\mathbf{X}}^{\xi}_{NW}(t)$ in terms of those operators for physically acceptable states. In fact, using the acting rules \eqref{eq:uio} one has that
\begin{equation}
\hat{X}^{j;\xi}_{NW}(t) = \underline{\hat{Q}^{j;\xi}_{phys}(\tau)} - \frac{\hat{\Pi}_{phys}^{j}}{\hat{\Pi}^{0}_{phys}}:\underline{\hat{Q}^{0;\xi}_{phys}(\tau)} + \frac{\hat{\Pi}_{phys}^{j}}{\hat{\Pi}^{0}_{phys}}t. \label{eq:HegUI}
\end{equation}
Relations \eqref{eq:RRRES} and \eqref{eq:HegUI} make explicit the fact that physically acceptable states $\psi$ belonging to the domain $D_{\hat{X}_{NW}^{\xi;j}(t)}$ must satisfy, among other restrictions, the condition $\psi \in D_{\underline{\hat{Q}^{\mu;\xi}_{phys}(\tau)}}$, even if $D_{\hat{X}_{NW}^{\xi;j}}$ authorize more general states. However, as presented in \cite{Taillebois2021a}, states that satisfy this condition cannot be strictly localized or localized with exponential tails with regard to the POVM associated with $\underline{\hat{Q}^{\mu;\xi}_{phys}(\tau)}$ and, consequently, the evolution with respect to proper-time $\tau$ is not subject to the violation of causality foreseen by Hegerfeldt. This fact allows us to question whether the observed violation for $\hat{X}_{NW}^{\xi;j}(t)$ does not correspond to an apparent phenomenon originated from the restriction process that replaces the classic intrinsic parameter $\tau$ with the effective parameter $t$.

The relationship \eqref{eq:HegUI} allows the properties of $\hat{X}^{j;\xi}_{NW}(t)$ to be interpreted in terms of the operators $\underline{\hat{Q}^{\mu;\xi}_{phys}(\tau)}$ and will be useful in understanding the origin of Hegerfeldt's paradox for $\hat{X}^{j;\xi}_{NW}(t)$. To do this, we use relation \eqref{eq:IdentP1} to rewrite \eqref{eq:HegUI} as
\begin{widetext}
\begin{equation}
\hat{X}^{j;\xi}_{NW}(t) =  \underline{\hat{Q}^{j;\xi}_{phys}(0)} - \sum_{l'=0}^{\infty}\sum_{m_{z}'=-l'}^{l'}\int_{\mathbb{R}}dt' t' \frac{\hat{\Pi}_{phys}^{3}}{\hat{\Pi}^{0}_{phys}}:\ket{\psi^{t'+t,l',m_{z}'}_{0;\xi}}\bra{\psi^{t'+t,l',m_{z}'}_{0;\xi}} \label{eq:poiu}
\end{equation}
\end{widetext}
This relation shows that the differences in the results observed when applying $\hat{X}^{j;\xi}_{NW}(t)$ over any state $\ket{\psi}$ for different values of $t$ is due exclusively to the second term to the right of \eqref{eq:poiu}, since the overlap of  $t' \frac{\hat{\Pi}_{phys}^{3}}{\hat{\Pi}^{0}_{phys}}$ contributions will be different for each value of $t$. This term is associated with the temporal uncertainty of the $\ket{\psi}$ state with respect to the $\underline{\hat{Q}^{0;\xi}_{phys}(\tau)}$ operator, which reveals that the violation of causality predicted by Hegerfeldt is due solely to the existence of this uncertainty.

Since the basis associated with the states $\ket{\psi^{t',l,m_{z}}_{\tau;\xi}}$ is improper and not orthogonal, the physical states belonging to a well-defined energy sign subspace will always be non-local in time. Consequently, measurements of $\hat{X}^{j;\xi}_{NW}(t)$ will always receive contributions associated with the description of the particle at different times. This characteristic suggests that Hegerfeldt's violation for $\hat{\mathbf{X}}_{NW}^{\xi}(t)$ may be an apparent phenomenon resulting from the temporal indeterminacy of the state. From this point of view, the apparent superluminal propagation would result from the fact that the particle can potentially be found in the past or in the future of the instant of time $t$ in which the observer defines the located state with respect to $\hat{X}^{j;\xi}_{NW}(t)$. Therefore, future observations of $\hat{X}^{j;\xi}_{NW}(t')$, with $t'>t$, could receive contributions of time instants $t''\neq t$ that could spread to distant regions without exceeding the speed of light, thus producing the apparent violation of causality.

\section{\label{sc:disc}Discussion}

In the present work it was shown that the single-particle formalism parametrized by the system's proper-time can be consistently connected to usual descriptions parametrized by classical observers parameters. This connection is obtained through a restriction operation that mixes contributions of different values of proper-time. Such a procedure allows to retrieve the concepts of Newton-Wigner position and Kijowski time of detection in an effective way, enabling to interpret the properties of such operators and the related phenomena through the proper-time parametrized formalism, where every physically acceptable single particle quantum-state has an inherent time uncertainty. It is worth emphasizing that the Kijowski operator $[\underline{\hat{Q}^{0,+}_{phys}(\tau)}]_{\underline{\hat{Q}^{3;+}_{phys}(\tau)} - z \mapsto 0}$ obtained here agrees with more recent descriptions of the detection time distributions \cite{Egusquiza2008}, thus supporting the proposed restriction prescription. 

The Newton-Wigner position operator leads to a notion of localization that allows strictly localized states to be defined, thus presenting the causality issues predicted by Hegerfeldt's Theorem. The causality violation is small for Newton-Wigner states \cite{Ruijsenaars1981}, however, due to the absence of any information concerning the time uncertainty of such states in the usual formalism, this imposes challenging difficulties for the interpretation of $\hat{X}^{j;\xi}_{NW}(t)$ as an adequate notion of localization. Using the proper-time framework, any single-particle states with a well defined energy sign possess a time-uncertainty, which opens the way for a possible understanding of the superluminal spreading of Newton-Wigner states as an apparent consequence of the effective character of $\hat{X}^{j;\xi}_{NW}(t)$. In this sense, measurements of $\hat{X}^{j;\xi}_{NW}(t)$ would receive contributions from states representing the particle at many different times and, since the time of the particle is potentially in the past of the time of measurement, the particle's propagation to distant regions would not necessarily imply superluminal velocities. In this sense, the origin of the apparent causality paradox would be intimately connected to the inappropriate association of the measurement time with the system time.

\vspace{\baselineskip}
\begin{acknowledgments}
We are thankful for the support provided by Brazilian agencies CAPES (PROCAD2013), CNPq (\#459339/2014-1, \#312723/2018-0), FAPEG (PRONEX \#201710267000503, PRONEM \#201710267000540) and the Instituto Nacional de Ciência e Tecnologia - Informação Quântica (INCT-IQ \#465469/2014-0).
\end{acknowledgments}

\appendix

\section{\label{app:1} Brief account of previous results}

This appendix presents a brief account concerning some results from \cite{Taillebois2021a} that are relevant to the developments of the present work.

The proper-time parametrized formalism allows to describe the system's four-position in a given proper-time instant $\tau$ by means of the closed but non-self-adjoint components of the four-vector operator $\underline{\hat{Q}^{\mu}_{phys}(\tau)}$. Denoting by $\sigma^{3}$ the usual third Pauli matrix, the acting rules associated to this operator are given by
\begin{widetext}
\begin{subequations} \label{eq:uio}
\begin{eqnarray}
\check{Q}_{phys}^{0}(\tau) & = & \sigma^{3}\frac{E_{\bs{\pi}}}{m}\left[\frac{i}{m}\left(\bs{\pi}\cdot\nabla_{\bs{\pi}} + \frac{3}{2}\right) + \tau\right], \\
\check{Q}_{phys}^{j}(\tau) & = & \sigma^{3}\left[i\left(\frac{\partial}{\partial \pi_{j}} + \frac{\pi^{j}}{m^2}\bs{\pi}\cdot\nabla_{\bs{\pi}} + \frac{3}{2}\frac{\pi^{j}}{m^2}\right) +\frac{\pi^{j}}{m}\tau\right].
\end{eqnarray}
\end{subequations}

The POVM associated with the time operator $\underline{\hat{Q}^{0;\xi}_{phys}(\tau)}$ is given by the set of positive operators
\begin{equation}
\hat{E}_{\tau;\pm}(t) = \sum_{l=0}^{\infty}\sum_{m_{z}=-l}^{l}\ket{\psi^{t,l,m_z}_{\tau;\pm}}\bra{\psi^{t,l,m_z}_{\tau;\pm}}, \nonumber
\end{equation}
with
\begin{equation}
\int_{-\infty}^{\infty}dt \hat{E}_{\tau;\pm}(t) = \hat{I}^{\pm}_{phys} \label{eq:IdentP1}
\end{equation}
and
\begin{equation}
\psi^{t,l,m_z}_{\tau;\pm}(\bs{\pi}) = \sqrt{\frac{m}{2\uppi}}\frac{Y^{l,m_{z}}(\Omega_{\pi})}{r_{\pi}^{3/2}}\hspace{-0.1cm}\left(\frac{r_{\pi}}{m}\right)^{im\tau}\hspace{-0.1cm}\left(\frac{r_{\pi}}{E_{r_{\pi}} + m}\right)^{\mp imt}. \nonumber
\end{equation}

The POVM associated with the proper-time parametrized position operator $\underline{\hat{Q}^{3;\xi}_{phys}(\tau)}$ is given by the set of positive operators
\begin{equation}
\hat{E}_{\tau;\xi}(z) = \sum_{m_{z}\in\mathbb{Z}}\int_{-\infty}^{-1/4}d\lambda\ket{\psi^{z,\lambda,m_{z}}_{\tau;\xi}}\bra{\psi^{z,\lambda,m_{z}}_{\tau;\xi}}, \nonumber
\end{equation}
with
\begin{equation}
\int_{\mathbb{R}}dz\hat{E}_{\tau;\xi}(z) = \hat{I}^{\xi}_{phys} \label{eq:IdentP2}
\end{equation}
and
\begin{equation}
\psi^{z,\lambda,m_{z}}_{\tau;\xi}(\bs{\pi}) = \frac{\sqrt{\sinh(\uppi\Lambda(\lambda))}}{m}\frac{|\Gamma(\frac{1}{2} + |m_{z}| + i\Lambda(\lambda))|}{(2\uppi)^{3/2}}\frac{e^{im\tau\ln(\sec\nu_{\pi})}e^{-im\xi z\nu_{\pi}}e^{im_{z}\varphi_{\pi}}}{(\sec\nu_{\pi})^{3/2}}P^{-|m_{z}|}_{-\frac{1}{2} + i\Lambda(\lambda)}(\cosh\omega_{\pi}). \nonumber
\end{equation}
\end{widetext}

\bibliography{bib/Bibliografia,bib/bibtiqr}

\end{document}